\documentclass{article}
\usepackage{amsmath}
\usepackage{graphicx}

\begin{document}

\title{From Quantum Field Theory\\
to Quantum Mechanics}
\author{Nuno Barros e S\'a$^1$ $^2$ \thanks{%
nuno.bf.sa@uac.pt} \and Cl\'audio Gomes$^1$ $^3$ \thanks{%
claudio.fv.gomes@uac.pt}}
\date{$^1$ DCFQE, Universidade dos A\c cores, 9500-801 Ponta Delgada,
Portugal\\
$^2$ OKEANOS, Universidade dos A\c cores, 9901-862 Horta, Portugal\\
$^3$ Centro de F\'isica das Universidades do Minho e do Porto, Rua do Campo
Alegre s/n, 4169-007 Porto, Portugal}
\maketitle

\begin{abstract}
We construct the algebra of operators acting on the Hilbert spaces of
Quantum Mechanics for systems of $N$ identical particles from the field
operators acting in the Fock space of Quantum Field Theory by providing the
explicit relation between the position and momentum operators acting in the
former spaces and the field operators acting on the latter. This is done in
the context of the non-interacting Klein-Gordon field. It may not be
possible to extend the procedure to interacting field theories since it
relies crucially on particle number conservation. We find it nevertheless
important that such an explicit relation can be found at least for free
fields. It also comes out that whatever statistics the field operators obey
(either commuting or anticommuting), the position and momentum operators
obey commutation relations. The construction of position operators raises
the issue of localizability of particles in Relativistic Quantum Mechanics,
as the position operator for a single particle turns out to be the
Newton-Wigner position operator. We make some clarifications on the
interpretation of Newton-Wigner localized states and we consider the
transformation properties of position operators under Lorentz
transformations, showing that they do not transform as tensors, rather in a
manner that preserves the canonical commutation relations.
\end{abstract}

\bigskip

Since its inception, Quantum Field Theory suffered from a number of
problems, perhaps most notably: it inherited the conceptual problems already
present in the interpretation of Quantum Mechanics; the theory is plagued
with infinities - despite the successes of renormalization, the situation is
unconfortable - ; and the theory is not "dynamical", i. e., it does not
provide for a picture of the spacial and temporal evolution of a system.

None of these problems have had up to now a satisfactory resolution.
Nevertheless, no one questions the validity of both theories, Quantum
Mechanics and Quantum Field Theory, as their successes have been so
overwhelming, and many people believe that the resolution of these problems
shall come someday. Even if these theories are not "correct", it is
legitimate to believe that they must be correct in "some limit", as their
validity has been solidly proven experimentaly in many domains.

Given the proximity between Quantum Mechanics in first and second
quantisation, it would be sensible too to find the "limit" when Quantum
Field Theory can be approximated by Quantum Machanics, that is, Quantum
Mechanics should be contained in Quantum Field Theory, because the former
deals with quantum systems with fixed number of particles while the latter
deals with arbitrary numbers of particles. While there are some methods that
allow one to recover single particle Quantum Mechanics from Quantum Field
theory (see \cite{pad} for an extensive review on propagator methods and
\cite{schwartz}, Chap. 33, for approximative methods in the context of
effective field theories), it would be interesting to construct an explicit
relation between the two. Such a task seems formidable for interacting field
theories.

Here we propose a method by which from a theory for a free quantum field
(the Klein-Gordon field, for simplicity) one can construct all the quantum
mechanical systems with fixed numbers of particles in first quantization. In
particular, we construct the algebra of operators in first quantisation
(that is, the $x$'s and $p$'s) from the field operators in second
quantisation. We find it rewarding that, at least for free theories, the
reduction of Quantum Field Theory to Quantum Mechanics can be done - and can
be done for any number of particles.

We should note that while completing this article we became aware of the
work of Pav\v{s}i\v{c} \cite{pavsic} where very similar conclusions were
withdrawn. However, we use a different approach to some of problems which
are tackled in both works, and made some improvements on the results.

We finish by showing that the position operators constructed are the
Newton-Wigner operators (which were an early attempt to build localised
states in Relativistic Quantum Mechanics \cite{nw}), that they do not
transform as components of tensors under Lorentz transformation but rather
in a manner that preserves the canonical commutation relations, and that
localized states are indeed present in relativistic field theories, though
they do not have an invariant meaning: like the position operators they are
frame dependent concepts.

\section{The Klein-Gordon field}

We make a brief review the classical Klein-Gordon field, which can also be
found in most textbooks. The Lagrangian is%
\begin{equation}
L=\frac{1}{2}\left( \eta ^{\mu \nu }\partial _{\mu }\phi \partial _{\nu
}\phi -m^{2}\phi ^{2}\right)
\end{equation}%
The momentum conjugate to $\phi $ is (we use a $\left( +,-,-,-\right) $
metric)%
\begin{equation}
\pi =\dot{\phi}
\end{equation}%
and one can solve the Klein-Gordon equation with the result%
\begin{eqnarray}
\phi \left( x,t\right) &=&\int \frac{d^{3}k}{\left( 2\pi \right) ^{3/2}}%
\sqrt{\frac{\hbar }{2\omega _{k}}}\left[ a_{k}e^{-\mathrm{i}\omega _{k}t+%
\mathrm{i}kx}+a_{k}^{+}e^{\mathrm{i}\omega _{k}t-\mathrm{i}kx}\right]
\label{fi} \\
\pi \left( x,t\right) &=&\int \frac{d^{3}k}{\left( 2\pi \right) ^{3/2}%
\mathrm{i}}\sqrt{\frac{\hbar \omega _{k}}{2}}\left[ a_{k}e^{-\mathrm{i}%
\omega _{k}t+\mathrm{i}kx}-a_{k}^{+}e^{\mathrm{i}\omega _{k}t-\mathrm{i}kx}%
\right]  \label{pi}
\end{eqnarray}%
where the $a_{k}$ are complex constants. Here%
\begin{equation}
\omega _{k}=\sqrt{k^{2}+m^{2}}
\end{equation}%
and both $x$ and $k$ are to be understood as 3-dimensional quantities, e.
g., $kx$ stands for $\vec{k}\cdot \vec{x}$. Energy and linear momentum are
given by%
\begin{eqnarray}
H &=&\int d^{3}x\frac{1}{2}\left( \pi ^{2}+\left\Vert \vec{\nabla}\phi
\right\Vert ^{2}+m^{2}\phi ^{2}\right) =\int d^{3}k\hbar \omega
_{k}a_{k}^{+}a_{k}  \label{ha} \\
\vec{P} &=&-\int d^{3}x\pi \vec{\nabla}\phi =\int d^{3}k\hbar \vec{k}%
a_{k}^{+}a_{k}  \label{pe}
\end{eqnarray}

Quantization is performed replacing the classical fields $\phi \left(
x,t\right) $ by operators and imposing the commutation relations%
\begin{eqnarray}
\left[ \phi \left( x,t\right) ,\phi \left( y,t\right) \right] &=&0
\label{cf1} \\
\left[ \phi \left( x,t\right) ,\pi \left( y,t\right) \right] &=&\mathrm{i}%
\hbar \delta \left( x-y\right)  \label{cf2} \\
\left[ \pi \left( x,t\right) ,\pi \left( y,t\right) \right] &=&0  \label{cf3}
\end{eqnarray}%
which translate to replacing the classical coefficients $a_{k}$ by operators
and imposing the commutation relations%
\begin{eqnarray}
\left[ a_{k},a_{q}\right] &=&0  \label{com1} \\
\left[ a_{k},a_{q}^{+}\right] &=&\delta \left( k-q\right)  \label{com2}
\end{eqnarray}

We should note that the normalization of the coefficients $a_{k}$ can be
chosen at will. Frequently the weighting factor in (\ref{fi}) and (\ref{pi})
is chosen to be $\left( 2\omega _{k}\right) ^{-1}$ rather than $\left(
2\omega _{k}\right) ^{-1/2}$ because $d^{3}k\left( 2\omega _{k}\right) ^{-1}$
is Lorentz invariant; then the commutator $\left[ a_{k},a_{q}^{+}\right] $
becomes $2\omega _{k}\delta \left( k-q\right) $ which is again Lorentz
invariant. This is the preferred normalization in many textbooks in Field
Theory (see, e. g., \cite{peskin} or \cite{itzy}). Here we chose a different
normalization, which we find more practical for our purpose of getting
Quantum Mechanics from Quantum Field Theory, as the former is constructed in
a language which is not manifestly covariant.

That said, we immediately face a problem of operator ordering since the two
alternative expressions for energy and linear momentum in (\ref{ha})-(\ref%
{pe}) are not equivalent, given the ambiguity in the ordering of operators.
For instance,%
\begin{equation}
\int d^{3}x\frac{1}{2}\left( \pi ^{2}+\left\Vert \vec{\nabla}\phi
\right\Vert ^{2}+m^{2}\phi ^{2}\right) =\int d^{3}x\hbar \omega _{k}\frac{1}{%
2}\left( a_{k}^{+}a_{k}+a_{k}a_{k}^{+}\right)
\end{equation}%
We shall adopt the right hand side of (\ref{ha})-(\ref{pe}) as our
definitions of energy and linear momentum, that is, normal ordering, thus
skipping the issue of zero-point infinities. In terms of the Klein-Gordon
field they are%
\begin{eqnarray}
H &=&\int d^{3}x\frac{1}{2}\left( \pi ^{2}+\left\Vert \vec{\nabla}\phi
\right\Vert ^{2}+m^{2}\phi ^{2}+\mathrm{i}\left[ \mathcal{W}\phi ,\pi \right]
\right)  \label{ha2} \\
\vec{P} &=&\int d^{3}x\frac{1}{2}\left( -\left\{ \vec{\nabla}\phi ,\pi
\right\} +\mathrm{i}\left( \vec{\nabla}\phi \mathcal{W}\phi +\vec{\nabla}\pi
\mathcal{W}^{-1}\pi \right) \right)  \label{pe2}
\end{eqnarray}%
Here the operator%
\begin{equation}
\mathcal{W}^{n}=\left( m^{2}-\nabla ^{2}\right) ^{n/2}
\end{equation}%
is non-local and can be understood in the sense of an infinite power series%
\begin{equation}
\mathcal{W}^{n}\phi =m^{n}\left[ \phi -\frac{n}{2}\frac{\nabla ^{2}\phi }{%
m^{2}}+\frac{1}{2!}\frac{n}{2}\left( \frac{n}{2}-1\right) \frac{\nabla
^{2}\left( \nabla ^{2}\phi \right) }{m^{4}}+\cdots \right]
\end{equation}%
or in the form%
\begin{equation}
\mathcal{W}^{n}\phi \left( x,t\right) =\int d^{3}yB\left( x-y\right) \phi
\left( y,t\right)
\end{equation}%
with%
\begin{equation}
B\left( x\right) =\int \frac{d^{3}k}{\left( 2\pi \right) ^{3/2}}\omega
_{k}^{n}e^{\mathrm{i}kx}
\end{equation}%
Clearly expressions (\ref{ha2})-(\ref{pe2}) coincide with the classical
limit (\ref{ha})-(\ref{pe}) for commuting fields.

We can also compute the number of particles operator%
\begin{equation}
N=\int d^{3}ka_{k}^{+}a_{k}=\frac{1}{\hbar }\int d^{3}x\frac{1}{2}\left(
\phi \mathcal{W}\phi +\pi \mathcal{W}^{-1}\pi +\mathrm{i}\left[ \phi ,\pi %
\right] \right)
\end{equation}%
Curiously, in the classical limit%
\begin{equation}
N=\frac{1}{\hbar }\int d^{3}x\frac{1}{2}\left( \phi \mathcal{W}\phi +\pi
\mathcal{W}^{-1}\pi \right)
\end{equation}%
which can be easily checked to be a constant of motion. There is therefore a
semiclassical sense of "number of particles", its semiclassical nature being
noticeable in the emergence of Planck's constant. However, this expression
involves the non-local operator $\mathcal{W}^{n}$, meaning that one cannot
define the "density of particles", the number of particles being a strictly
global concept.

The states that describe $N$ identical particles with sharply defined momenta%
\begin{equation}
\left\vert k_{1}\cdots k_{N}\right\rangle =a_{k_{1}}^{+}\cdots
a_{k_{N}}^{+}\left\vert 0\right\rangle  \label{foc}
\end{equation}%
form a basis for the Hilbert space $\mathcal{H}_{S}^{N}$ of Quantum
Mechanics for a system with $N$ identical particles. The complete space of
states for the field theory is the direct sum
\begin{equation}
\mathcal{H}_{S}=\oplus _{N}\mathcal{H}_{S}^{N}
\end{equation}%
It is our purpose to show that it is possible, within each $\mathcal{H}%
_{S}^{N}$, to reconstruct the usual canonical variables of Quantum Mechanics
out of the field operators $\phi $ and $\pi $. However, since we are dealing
with identical particles, the operators that act on $\mathcal{H}_{S}^{N}$
cannot be usual canonical variables $X_{i}$ and $P_{i}$, since they are not
permutation invariant. We shall therefore first revisit the spaces $\mathcal{%
H}_{S}^{N}$.

\section{Systems of $N$ identical particles}

The states (\ref{foc}) can be constructed without the use of creation
operators from the direct products of one-particle basis states for each
particle%
\begin{equation}
\left\vert k_{1}\right\rangle \times \cdots \times \left\vert
k_{N}\right\rangle  \label{sa0}
\end{equation}%
by summing over permutations%
\begin{equation}
\left\vert k_{1}\cdots k_{N}\right\rangle _{S}=\frac{1}{\sqrt{N!}}%
\sum_{P}\left\vert k_{P\left( 1\right) }\right\rangle \times \cdots \times
\left\vert k_{P\left( N\right) }\right\rangle  \label{sa1}
\end{equation}%
Here we introduced the index $S$ to remind that the creation and
annihilation operators obey commutation relations (\ref{com1})-(\ref{com2}).
Had we used instead anticommutation relations%
\begin{eqnarray}
\left\{ a_{k},a_{q}\right\} &=&0  \label{cg1} \\
\left\{ a_{k}^{+},a_{q}\right\} &=&\delta \left( k-q\right)  \label{cg2}
\end{eqnarray}%
and the order $p$ of the permutation should have been taken into
consideration,%
\begin{equation}
\left\vert k_{1}\cdots k_{N}\right\rangle _{A}=\frac{1}{\sqrt{N!}}%
\sum_{P}\left( -1\right) ^{p}\left\vert k_{P\left( 1\right) }\right\rangle
\times \cdots \times \left\vert k_{P\left( N\right) }\right\rangle
\label{sa2}
\end{equation}%
This would be the space $\mathcal{H}_{A}^{N}$ of antisymmetric
wavefunctions. Both $\mathcal{H}_{S}^{N}$ and $\mathcal{H}_{A}^{N}$ are
subspaces of the space $\mathcal{H}^{N}$ generated by the kets (\ref{sa0}).

The only operators that map states in $\mathcal{H}_{S}^{N}$ into states in $%
\mathcal{H}_{S}^{N}$ and states in $\mathcal{H}_{A}^{N}$ into states in $%
\mathcal{H}_{A}^{N}$ are permutation invariant operators, which can be
constructed from any operator $O\left( X_{1},P_{1},\ldots
,X_{N},P_{N}\right) $ by summing over permutations%
\begin{equation}
O_{S}=\frac{1}{\sqrt{N!}}\sum_{P}O\left( X_{P\left( 1\right) },P_{P\left(
1\right) },\ldots ,X_{P\left( N\right) },P_{P\left( N\right) }\right)
\end{equation}%
We shall show that such operators can also be written using creation and
annihilation operators.

For that purpose we start by reminding that any polynomial operator in a
canonical pair, of order $m$ in $X$ and $n$ in $P$, can be written as a sum
of monomials with $X$ ordered to the left of the $P$. That happens because,
using the commutation relations $\left[ X,P\right] =\mathrm{i}\hbar $, one
can always write a monomial of order $m$ in $X$ and $n$ in $P$ in the form
\begin{equation}
X^{m}P^{n}+\sum_{k=1}^{\min \left( m,n\right) }\alpha _{i}\left( -\mathrm{i}%
\hbar \right) ^{k}X^{m-k}P^{n-k}
\end{equation}%
For example%
\begin{equation}
P^{3}XP^{2}X^{2}=X^{3}P^{5}+13\left( -\mathrm{i}\hbar \right)
X^{2}P^{4}+44\left( -\mathrm{i}\hbar \right) ^{2}XP^{3}+36\left( -\mathrm{i}%
\hbar \right) ^{3}P^{2}
\end{equation}%
For $N$ particles this result generalizes to the statement that any
polynomial operator of order $m_{i}$ in $X_{i}$ and $n_{i}$ in $P_{i}$ can
be written as a sum of monomials with $X_{i}$ ordered to the left of the $%
P_{i}$%
\begin{equation}
X_{1}^{m_{1}}P_{1}^{n_{1}}\ldots X_{N}^{m_{N}}P_{N}^{n_{N}}
\end{equation}%
Therefore all polynomial operators for identical particles can be written as
sums of parcels of the type%
\begin{equation}
\frac{1}{\sqrt{N!}}\sum_{P}X_{P\left( 1\right) }^{m_{1}}P_{P\left( 1\right)
}^{n_{1}}\ldots X_{P\left( N\right) }^{m_{N}}P_{P\left( N\right) }^{n_{N}}
\end{equation}

Now we note that sums over permutations can always be written as
combinations of ordinary sums. We start by writing%
\begin{eqnarray}
&&\sum_{P}O_{P\left( 1\right) }O_{p\left( 2\right) }\ldots O_{P\left(
N\right) }=  \notag \\
&=&\sum_{i_{1}}^{N}O_{i_{1}}\sum_{\substack{ i_{2}  \\ \left( i_{2}\neq
i_{1}\right) }}^{N}O_{i_{2}}\sum_{\substack{ i_{3}  \\ \left( i_{3}\neq
i_{1}\right)  \\ \left( i_{3}\neq i_{2}\right) }}^{N}O_{i_{3}}\ldots \sum
_{\substack{ i_{N}  \\ \left( i_{N}\neq i_{1}\right)  \\ \left( i_{N}\neq
i_{1}\right)  \\ \cdots  \\ \left( i_{N}\neq i_{N-1}\right) }}^{N}O_{i_{N}}
\label{au}
\end{eqnarray}%
And use the identity%
\begin{equation}
\sum_{a}\sum_{\substack{ b  \\ \left( b\neq a\right) }}f\left( a,b\right)
=\sum_{a}\sum_{b}f\left( a,b\right) -\sum_{a}\sum_{a}f\left( a,a\right)
\end{equation}%
For $N=2$ in (\ref{au}) this produces (omitting the sum signs, for
convenience)%
\begin{eqnarray}
&&\sum_{P}O_{P\left( 1\right) }O_{P\left( 2\right) }=O_{1}O_{2}+O_{2}O_{1}=
\notag \\
&=&O_{i}O_{\substack{ j  \\ \left( j\neq i\right) }}=O_{i}O_{j}-O_{i}O_{i}=
\end{eqnarray}%
For $N=3$, eq. (\ref{au}) can be used successively to get%
\begin{eqnarray}
&&\sum_{P}O_{P\left( 1\right) }O_{P\left( 2\right) }O_{P\left( 3\right) }=
\notag \\
&=&O_{1}O_{2}O_{3}+O_{1}O_{3}O_{2}+O_{2}O_{1}O_{3}+O_{2}O_{3}O_{1}+O_{3}O_{1}O_{2}+O_{3}O_{2}O_{1}=
\notag \\
&=&O_{i}O_{\substack{ j  \\ \left( j\neq i\right) }}O_{\substack{ k  \\ %
\left( k\neq i\right)  \\ \left( k\neq j\right) }}=O_{i}O_{\substack{ j  \\ %
\left( j\neq i\right) }}O_{\substack{ k  \\ \left( k\neq i\right) }}-O_{i}O
_{\substack{ j  \\ \left( j\neq i\right) }}O_{\substack{ j  \\ \left( j\neq
i\right) }}=  \notag \\
&=&O_{i}O_{j}O_{\substack{ k  \\ \left( k\neq i\right) }}-O_{i}O_{i}O
_{\substack{ k  \\ \left( k\neq i\right) }}-O_{i}O_{\substack{ j  \\ \left(
j\neq i\right) }}O_{\substack{ j  \\ \left( j\neq i\right) }}=  \notag \\
&=&O_{i}O_{j}O_{k}-O_{i}O_{j}O_{i}-O_{i}O_{i}O_{k}+O_{i}O_{i}O_{i}-O_{i}O_{j}O_{j}+O_{i}O_{i}O_{i}=
\notag \\
&=&O_{i}O_{j}O_{k}-O_{i}O_{j}O_{j}-O_{j}O_{i}O_{j}-O_{j}O_{j}O_{i}+2O_{i}O_{i}O_{i}
\end{eqnarray}%
And so on for higher $N$.

Hence, all polynomial operators for identical particles can be obtained from
parcels of the type%
\begin{equation}
\sum_{i_{1}=1}^{N}\ldots
\sum_{i_{N}=1}^{N}X_{i_{1}}^{m_{1}}P_{i1}^{n_{1}}\ldots
X_{i_{N}}^{m_{N}}P_{i_{N}}^{n_{N}}=%
\sum_{i_{1}=1}^{N}X_{i_{1}}^{m_{1}}P_{i_{1}}^{n_{1}}\ldots
\sum_{i_{N}=1}^{N}X_{i_{N}}^{m_{N}}P_{i_{N}}^{n_{N}}
\end{equation}%
or with a lesser number of sums. These parcels, in turn, can be obtained
from the building blocks%
\begin{equation}
\sum_{i=1}^{N}X_{i}^{m}P_{i}^{n}  \label{glu}
\end{equation}%
by multiplication. As for the parcels with a lesser number of sums, they
again can be written using these building blocks, as it is evident by the
following example%
\begin{equation}
\sum_{i=1}^{N}X_{i}P_{i}X_{i}P_{i}=\sum_{i=1}^{N}X_{i}^{2}P_{i}^{2}-i\hbar
\sum_{i=1}^{N}X_{i}P_{i}
\end{equation}%
In conclusion, if we find a representation for the operators (\ref{glu}), we
can construct all permutation invariant operators in either $\mathcal{H}%
_{S}^{N}$ or $\mathcal{H}_{A}^{N}$ by summing and multiplying them.

\section{Fock space}

Indeed we can see that the operators of the type (\ref{glu}) can be written
in the form%
\begin{equation}
\sum_{i=1}^{N}X_{i}^{m}P_{i}^{n}=\int \frac{dqdxdk}{2\pi }x^{m}\left( \hbar
k\right) ^{n}\exp \left[ \mathrm{i}\left( k-q\right) x\right] a_{q}^{+}a_{k}
\label{quatro}
\end{equation}%
independently of $N$ and independently of whether commutation or
anticommutation relations are used. This can be done by computing the
spectra of the operators on the left and on the right sides of (\ref{quatro}%
) and checking that they match for all basis states using (\ref{sa1}) or (%
\ref{sa2}) on the left hand side and (\ref{foc}) on the right hand side. For
example, for two-particle states%
\begin{equation}
\left\vert k_{1}k_{2}\right\rangle =\frac{1}{\sqrt{2}}\left( \left\vert
k_{1}\right\rangle \times \left\vert k_{2}\right\rangle \pm \left\vert
k_{2}\right\rangle \times \left\vert k_{1}\right\rangle \right)
=a_{k_{1}}^{+}a_{k_{2}}\left\vert 0\right\rangle
\end{equation}%
and it is easy to check that%
\begin{eqnarray}
&&\frac{1}{\sqrt{2}}\left( \left\langle q_{1}\right\vert \times \left\langle
q_{2}\right\vert \pm \left\langle q_{2}\right\vert \times \left\langle
q_{1}\right\vert \right) \left[ X_{1}^{m}P_{1}^{n}+X_{2}^{m}P_{2}^{n}\right]
\times  \notag \\
&&\times \frac{1}{\sqrt{2}}\left( \left\vert k_{1}\right\rangle \times
\left\vert k_{2}\right\rangle \pm \left\vert k_{2}\right\rangle \times
\left\vert k_{1}\right\rangle \right)  \notag \\
&=&\left\langle q_{1}q_{2}\right\vert \int \frac{dqdxdk}{2\pi }%
q^{l}x^{m}\left( \hbar k\right) ^{n}\exp \left[ \mathrm{i}\left( k-q\right) x%
\right] a_{q}^{+}a_{k}\left\vert k_{1}k_{2}\right\rangle =  \notag \\
&=&\int \frac{dx}{2\pi }x^{m}\left[ k_{1}^{n}e^{\mathrm{i}\left(
k_{1}-q_{1}\right) x}\delta \left( q_{2}-k_{2}\right) \pm k_{1}^{n}e^{%
\mathrm{i}\left( k_{1}-q_{2}\right) x}\delta \left( q_{1}-k_{2}\right) \pm
\right.  \notag \\
&&\times \left. k\pm k_{2}^{n}e^{\mathrm{i}\left( k_{2}-q_{1}\right)
x}\delta \left( q_{2}-k_{1}\right) +k_{2}^{n}e^{\mathrm{i}\left(
k_{2}-q_{2}\right) x}\delta \left( q_{1}-k_{1}\right) \right] \hbar ^{n}
\end{eqnarray}%
Notice that the result is valid irrespectively of whether one uses
commutation or anticommutation relations, that is, either in $\mathcal{H}%
_{S}^{2}$ or in $\mathcal{H}_{A}^{2}$. The generalization to higher $N$ is
straightforward.

The cross terms between $\mathcal{H}^{M}$ and $\mathcal{H}^{N}$ with $M\neq
N $ cannot be computed for the operator on the left hand side of (\ref%
{quatro}) but for the operator on right hand side they can, and they vanish
identically. One should therefore look upon the right hand side of (\ref%
{quatro}) as an extension of its left hand side which is valid for the whole
of Fock space.

In three-dimensional space it is easy to generalize the building blocks (\ref%
{quatro}) to%
\begin{eqnarray}
&&%
\sum_{i=1}^{N}X_{i}^{m_{x}}Y_{i}^{m_{y}}Z_{i}^{m_{z}}P_{xi}^{n_{x}}P_{yi}^{n_{y}}P_{zi}^{n_{z}}=\int
\frac{d^{3}qd^{3}xd^{3}k}{\left( 2\pi \right) ^{3}}\times  \notag \\
&&\times x^{m_{z}}y^{m_{y}}z_{x}^{m_{z}}\left( \hbar k_{x}\right)
^{n_{x}}\left( \hbar k_{y}\right) ^{n_{y}}\left( \hbar k_{z}\right)
^{n_{z}}\exp \left[ \mathrm{i}\left( \vec{k}-\vec{q}\right) \cdot \vec{x}%
\right] a_{\vec{q}}^{+}a_{\vec{k}}  \label{cinco}
\end{eqnarray}

Let us have a look at the expressions for the operators "total momentum" and
"sum of positions"
\begin{equation}
\vec{P}=\sum_{i=1}^{N}\vec{P}_{i}\qquad \text{and\qquad }\vec{X}%
=\sum_{i=1}^{N}\vec{X}_{i}
\end{equation}%
derived using (\ref{cinco})%
\begin{eqnarray}
\vec{P} &=&\hbar \int \frac{d^{3}qd^{3}xd^{3}k}{\left( 2\pi \right) ^{3}}%
\vec{k}e^{\mathrm{i}\left( k-q\right) x}a_{q}^{+}a_{k}=\hbar \int d^{3}k\vec{%
k}a_{\vec{k}}^{+}a_{\vec{k}}  \label{xe} \\
\vec{X} &=&\int \frac{d^{3}qd^{3}xd^{3}k}{\left( 2\pi \right) ^{3}}\vec{x}e^{%
\mathrm{i}\left( k-q\right) x}a_{q}^{+}a_{k}=  \notag \\
&=&\mathrm{i}\int d^{3}ka_{k}^{+}\frac{\partial a_{k}}{\partial \vec{k}}=-%
\mathrm{i}\int d^{3}k\frac{\partial a_{k}^{+}}{\partial \vec{k}}a_{k}
\label{pes}
\end{eqnarray}%
One can compute the commutator of this two operators%
\begin{equation}
\left[ X_{\alpha },P_{\beta }\right] =\mathrm{i}\hbar \int
d^{3}pd^{3}kk_{\beta }\left[ a_{p}^{+}\frac{\partial a_{p}}{\partial
p_{\alpha }},a_{\vec{k}}^{+}a_{\vec{k}}\right] =\mathrm{i}\hbar N\delta
_{\alpha \beta }  \label{com}
\end{equation}%
where we have used the identity%
\begin{equation}
\left[ AB,CD\right] =A\left[ B,C\right] _{\pm }D+\left[ A,C\right] _{\pm
}DB-AC\left[ D,B\right] _{\pm }-C\left[ D,A\right] _{\pm }B
\end{equation}%
which is valid for for commutators ($-$ sign) and anticommutators ($+$
sign). The commutator (\ref{com}) gives the expected result for both types
of statistics, being proportional to the number of particles operator.

\section{Creation and anihilation of particles at fixed points in space}

The Fourier transform of the set of creation operators for particles with
fixed values of momentum%
\begin{equation}
a_{x}^{+}=\int \frac{d^{3}k}{\left( 2\pi \right) ^{3/2}}e^{-ikx}a_{k}^{+}
\label{opx}
\end{equation}%
can be seen as a set of creation operators for particles at fixed positions $%
x$ since its action on the vacuum produces precisely what is interpreted in
Quantum Mechanics as a localized state at position $x$. Multiparticle states
of $N$ localized particles can be constructed from these operators in the
same manner as it is done with states of $N$ particles with definite momenta%
\begin{equation}
\left\vert x_{1}\cdots x_{N}\right\rangle =a_{x_{1}}^{+}\cdots
a_{x_{N}}^{+}\left\vert 0\right\rangle
\end{equation}%
The position creation operators obey similar commutation or anticommutation
relations to the momentum creation operators,%
\begin{eqnarray}
\left[ a_{x},a_{y}\right] _{\pm } &=&0 \\
\left[ a_{x}^{+},a_{y}\right] _{\pm } &=&\delta \left( x-y\right)
\end{eqnarray}%
The inverse relations are%
\begin{equation}
a_{k}^{+}=\int \frac{d^{3}x}{\left( 2\pi \right) ^{3/2}}e^{\mathrm{i}%
kx}a_{x}^{+}  \label{opp}
\end{equation}

The building blocks (\ref{cinco}) become%
\begin{eqnarray}
&&%
\sum_{i=1}^{N}X_{i}^{m_{x}}Y_{i}^{m_{y}}Z_{i}^{m_{z}}P_{xi}^{n_{x}}P_{yi}^{n_{y}}P_{zi}^{n_{z}}=\int
\frac{d^{3}sd^{3}rd^{3}k}{\left( 2\pi \right) ^{3}}\times  \notag \\
&&\times r_{x}^{m_{z}}r_{y}^{m_{y}}r_{z}^{m_{z}}\left( \hbar k_{x}\right)
^{n_{x}}\left( \hbar k_{y}\right) ^{n_{y}}\left( \hbar k_{z}\right)
^{n_{z}}\exp \left[ \mathrm{i}\left( \vec{r}-\vec{s}\right) \cdot \vec{k}%
\right] a_{r}^{+}a_{s}
\end{eqnarray}%
Using position creation operators, the operators (\ref{xe})-(\ref{pes})
become
\begin{eqnarray}
\vec{X} &=&\int d^{3}x\vec{x}a_{\vec{x}}^{+}a_{\vec{x}} \\
\vec{P} &=&-\mathrm{i}\int d^{3}xa_{x}^{+}\frac{\partial a_{x}}{\partial x}=%
\mathrm{i}\int d^{3}x\frac{\partial a_{x}^{+}}{\partial x}a_{x}
\end{eqnarray}%
We see that a complete symmetry is provided by the Fourier transform which
allows one to use either position or momentum creation and annihilation
operators.

The following conclusions can be withdrawn from these results:

- It is possible to construct all operators involving permutation invariant
combinations of the canonical variables which act in each of the $\mathcal{H}%
_{S}^{N}$ and $\mathcal{H}_{A}^{N}$ spaces for all $N$ using the set of
creation and annihilation operators obeying respectively commutation and
anticommutation relations

- Whichever statistics is used for the creation and annihilation operators,
the resulting operators obey commutation relations.

- These operators are extensions to the whole Fock space. They map $\mathcal{%
H}^{N}$ onto $\mathcal{H}^{N}$ and therefore have vanishing cross products
for different values of $N$.

- Using the transformations (\ref{opx}) and (\ref{opp}) one can freely
switch between the momentum and the position representations and describe
the Hilbert state either in terms of basis states of particles with "well
defined momenta" or with "well defined positions".

\section{Position operator in Quantum Field Theory}

As we have seen, the set of creation and annihilation operators allows us to
construct all states in Fock space and all the relevant operators for
identical particles, whichever the statistics that they obey. It is
interesting to go back to the Klein-Gordon field and see what do these
operators look like when expressed in terms of the fields $\phi \left(
x,t\right) $ and $\pi \left( x,t\right) $ using the inverse relations to (%
\ref{fi})-(\ref{pi})%
\begin{equation}
a_{k}=\int \frac{d^{3}x}{\left( 2\pi \right) ^{3/2}}\left[ \sqrt{\frac{%
\omega _{k}}{2\hbar }}\phi \left( x,t\right) +\frac{\mathrm{i}}{\sqrt{2\hbar
\omega _{k}}}\pi \left( x,t\right) \right] e^{\mathrm{i}\omega _{k}t-\mathrm{%
i}kx}
\end{equation}%
Doing so for the "sum of momenta" (\ref{pes}), or total momentum, one gets
back to (\ref{pe2}). And for the "sum of positions" (\ref{xe}), one gets%
\begin{equation}
\vec{X}=\frac{1}{\hbar }\int d^{3}x\vec{x}\frac{1}{2}\left( \left( \mathcal{W%
}^{1/2}\phi \right) ^{2}+\left( \mathcal{W}^{-1/2}\pi \right) ^{2}+\mathrm{i}%
\left[ \mathcal{W}^{1/2}\phi ,\mathcal{W}^{-1/2}\pi \right] \right)
\end{equation}%
Again we see that in the classical limit one gets a non-local operator%
\begin{equation*}
\vec{X}=\frac{1}{\hbar }\int dV\frac{1}{2}\left( \left( \mathcal{W}%
^{1/2}\phi \right) ^{2}+\left( \mathcal{W}^{-1/2}\pi \right) ^{2}\right)
\vec{x}
\end{equation*}

We can use the Hamiltonian (\ref{ha}) to compute the time evolution of
operators, with the result%
\begin{eqnarray}
\frac{dN}{dt} &=&\frac{1}{\mathrm{i}\hbar }\left[ N,H\right] =0 \\
\frac{d\vec{P}}{dt} &=&\frac{1}{\mathrm{i}\hbar }\left[ \vec{P},H\right] =0
\\
\frac{d\vec{X}}{dt} &=&\frac{1}{\mathrm{i}\hbar }\left[ \vec{X},H\right]
=\int d^{3}k\frac{\vec{k}}{\omega _{k}}a_{k}^{+}a_{k}
\end{eqnarray}%
Going back to our identification of operators (\ref{cinco}), the last
operator can be identified with%
\begin{equation*}
\int d^{3}k\frac{\vec{k}}{\omega _{k}}a_{k}^{+}a_{k}=\sum_{i}\frac{P_{i}}{%
\sqrt{\left( m\hbar \right) ^{2}+P_{i}^{2}}}
\end{equation*}%
as it should for a collection of non-interacting relativistic particles.

We should also remark that commutation relations (\ref{com1})-(\ref{com2})
for the $a$'s are compatible with the canonical relations (\ref{cf1})-(\ref%
{cf2})-(\ref{cf3}) for the fields. Anticommutation relations (\ref{cg1})-(%
\ref{cg2}) for the $a$'s would imply the following anticommutation rules for
the fields%
\begin{eqnarray}
\left\{ \phi \left( x,t\right) ,\phi \left( y,t\right) \right\}  &=&\hbar
\int \frac{d^{3}k}{\left( 2\pi \right) ^{3}}\frac{1}{\omega _{k}}e^{\mathrm{i%
}k\left( x-y\right) }=\hbar \mathcal{W}^{-1}\delta \left( x-y\right)  \\
\left\{ \phi \left( x,t\right) ,\pi \left( y,t\right) \right\}  &=&0 \\
\left\{ \pi \left( x,t\right) ,\pi \left( y,t\right) \right\}  &=&\hbar \int
\frac{d^{3}k}{\left( 2\pi \right) ^{3}}\omega _{k}e^{\mathrm{i}k\left(
x-y\right) }=\hbar \mathcal{W}\delta \left( x-y\right)
\end{eqnarray}%
These functions do not vanish for spacelike separations and this is
precisely one of the features showing up in the spin-statistics theorem \cite%
{pauli} (see \cite{spin} for a complete review on this theorem) which
excludes Fermi-Dirac statistics for integer spin fields, as is the case of
the Klein-Gordon field. We included here the case of anticommuting variables
because they can be used for semi-integer spin fields and their treatment is
analogous to the one performed here.

\section{Behaviour of operators under Lorentz transformations}

\begin{figure}
\centering
\includegraphics[width=.75\textwidth]{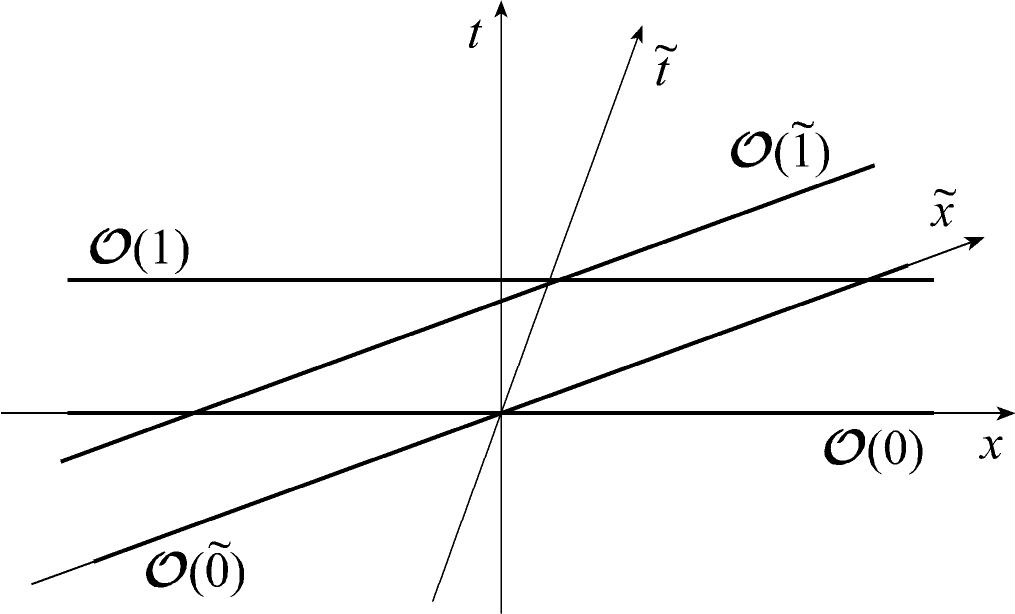}
\caption{Operators constructed by integrating local fields over spacelike surfaces cannot in general have simple transformation properties under Lorentz transformations since $\mathcal{O}(0)$ transforms to $\tilde{\mathcal{O}}(0)$ and not to $\tilde{\mathcal{O}}(\tilde{0})$, which is the transform of $\mathcal{O}(\tilde{0})$. That will happen only if $\mathcal{O}(0)=\mathcal{O}(\tilde{0})$}
\label{fig}
\end{figure}

Classically one can use the energy-momentum tensor $T^{\mu \nu }$ to
construct the momentum four-vector%
\begin{equation}
P^{\nu }=\int dS_{\mu }T^{\mu \nu }
\end{equation}%
where the integration is done on a space-like three-surface. At first sight
such a quantity should depend on the specific surface on which it is
calculated (fig. \ref{fig}). In particular, for a given Lorentz frame, on
each constant time three surface (that is, on a space volume for a given
time) this four vector should have a different value $P^{\mu }\left(
t\right) $, which in turn should be different from any constant time three
surface on another Lorentz $P^{\mu }\left( t^{\prime }\right) $. It is the
fact that $T^{\mu \nu }$ satisfies a conservation law%
\begin{equation}
\partial _{\mu }T^{\mu \nu }=0
\end{equation}%
that guarantees that $P^{\mu }\left( t_{1}\right) =P^{\mu }\left(
t_{2}\right) $ for $t_{1}\neq t_{2}$ and that $P^{\mu }\left( t\right)
=P^{\mu }\left( t^{\prime }\right) $ (see \cite{gold}).

That $P^{\mu }\left( t\right) $ does not depend on $t$ and that it behaves
like a four vector under Lorentz transformations can be seen easily using
the creation and annihilation operators formalism where one can write the
field in the form (\ref{fi}). Since this expression is valid in one frame $%
x^{\mu }$, it must be valid in another frame%
\begin{equation*}
\tilde{x}^{\mu }=\Lambda _{\;\nu }^{\mu }x^{\nu }
\end{equation*}%
for some other operators $\tilde{a}_{k}$
\begin{eqnarray}
\phi \left( x^{\mu }\right) &=&\int \frac{d^{3}k}{\left( 2\pi \right) ^{3/2}}%
\sqrt{\frac{\hbar }{2\omega _{k}}}\left[ a_{k}e^{-\mathrm{i}k_{\mu }x^{\mu
}}+a_{k}^{+}e^{\mathrm{i}k_{\mu }x^{\mu }}\right] =  \notag \\
&=&\int \frac{d^{3}k}{\left( 2\pi \right) ^{3/2}}\sqrt{\frac{\hbar }{2\omega
_{k}}}\left[ \tilde{a}_{k}e^{-\mathrm{i}k_{\mu }\tilde{x}^{\mu }}+\tilde{a}%
_{k}^{+}e^{\mathrm{i}k_{\mu }\tilde{x}^{\mu }}\right] =\phi \left( \tilde{x}%
^{\mu }\right)
\end{eqnarray}%
Equality of the two expressions implies that%
\begin{equation}
\sqrt{\omega _{k}}\tilde{a}_{k}=\sqrt{\omega _{\tilde{k}}}a_{\tilde{k}}
\end{equation}%
where $\tilde{k}_{\nu }=k_{\mu }\Lambda _{\;\nu }^{\mu }$. Had we used the
covariant normalization suggested after eqs. (\ref{com1})-(\ref{com2}) and
this transformation law would have had a simpler form, with no $\sqrt{\omega
}$ factors. Now we have from (\ref{ha})-(\ref{pe})%
\begin{eqnarray}
\tilde{P}^{\mu } &=&\int d^{3}k\hbar k^{\mu }\tilde{a}_{k}^{+}\tilde{a}%
_{k}=\int d^{3}\tilde{k}\frac{\omega _{k}}{\omega _{\tilde{k}}}\hbar k^{\mu }%
\sqrt{\frac{\omega _{\tilde{k}}}{\omega _{k}}}a_{\tilde{k}}^{+}\sqrt{\frac{%
\omega _{\tilde{k}}}{\omega _{k}}}a_{\tilde{k}}=  \notag \\
&=&\int d^{3}\tilde{k}\hbar k^{\mu }a_{\tilde{k}}^{+}a_{\tilde{k}}=\int d^{3}%
\tilde{k}\hbar \tilde{k}^{\nu }a_{\tilde{k}}^{+}a_{\tilde{k}}^{\nu }\left(
\Lambda ^{-1}\right) _{\nu }^{\;\mu }=P^{\nu }\left( \Lambda ^{-1}\right)
_{\nu }^{\;\mu }  \label{pro1}
\end{eqnarray}

Following the same procedure, the number operator is shown to behave as a
scalar%
\begin{equation}
\tilde{N}=N
\end{equation}%
However, no such type of transformation should be expected for the position
operators $X^{i}\left( t\right) $, and indeed for any operator that is not
constant in time. Of course any operator constructed for $\tilde{t}$
constant can be expressed in terms of operators constructed for $t$ constant
but the transformation law shall typically be complicated, e. g.,%
\begin{equation}
\tilde{X}^{i}\left( \tilde{0}\right) =i\int d^{3}k\tilde{a}_{k}^{+}\frac{%
\partial \tilde{a}_{k}}{\partial k^{i}}=i\int d^{3}\tilde{k}\sqrt{\frac{%
\omega _{k}}{\omega _{\tilde{k}}}}a_{\tilde{k}}^{+}\frac{\partial \tilde{k}%
^{j}}{\partial k^{i}}\frac{\partial }{\partial \tilde{k}^{j}}\left( \sqrt{%
\frac{\omega _{\tilde{k}}}{\omega _{k}}}a_{\tilde{k}}\right)  \label{pro2}
\end{equation}%
The fact that the position operators do not behave under Lorentz
transformations as tensors should not be seen as a drawback, but rather as
what should be expected: they are frame-dependent concepts and are
constructed so that they obey canonical commutation relations with the
momentum operators, can be verified from (\ref{pro1}) and (\ref{pro2})%
\begin{equation}
\left[ \tilde{X}^{i}\left( \tilde{0}\right) ,\tilde{P}^{j}\right] =\mathrm{i}%
\hbar N\delta ^{ij}=\mathrm{i}\hbar \tilde{N}\delta ^{ij}
\end{equation}%
In fact, the canonical commutation relations would not be covariant if the
position operators transformed like tensors.

However, for rotations $\omega _{\tilde{k}}=\omega _{k}$, $\tilde{t}=t$ and,
since $\tilde{k}^{j}$ depends linearly on $k^{i}$,%
\begin{equation}
\frac{\partial \tilde{k}^{j}}{\partial k^{i}}=R_{\;j}^{i}
\end{equation}%
where $R_{\;j}^{i}$ is a rotation matrix. Hence, eq. (\ref{pro2}) becomes%
\begin{equation}
\tilde{X}^{i}\left( \tilde{0}\right) =R_{\;j}^{i}\mathrm{i}\int d^{3}\tilde{k%
}a_{\tilde{k}}^{+}\frac{\partial a_{\tilde{k}}}{\partial \tilde{k}^{j}}%
=R_{\;j}^{i}X^{j}\left( 0\right)
\end{equation}

\section{Localization in Relativistic Quantum Mechanics}

It is the case that one can indeed construct position operators and find
states with sharply defined values for the position for the Klein-Gordon
field, in contrast with a common saying that "it is not possible to localize
a relativistic particle". One should therefore clarify this point. In a
famous article \cite{nw} Newton and Wigner studied the concept of
localization in Quantum Mechanics while attempting to satisfy the requisites
of Special Relativity too. Their work, togheter with the important work of
Foldy wnd Wouthuysen \cite{nwa} and \cite{nwb} generated some discussion on
the literature regarding position operators in Relativistic Quantum
Mechanics and their physical meaning lasting to these days \cite%
{pavsic,nw1,nw2,nw3,nw4,nw5}.

Newton and Wigner worked outside the context of Quantum Field Theory, rather
they searched for one-particle localized relativistic wavefunctions,
starting from momentum space $\psi \left( p\right) $ where an inner product
can be constructed using a relativistic invariant measure%
\begin{equation}
\left\langle \varphi |\psi \right\rangle =\int \frac{d^{3}p}{\omega _{p}}%
\varphi ^{\ast }\left( p\right) \psi \left( p\right)  \label{oito1}
\end{equation}%
And they assumed that the wavefunction in position space should be given by
the Fourier transform of $\psi \left( p\right) $ computed using the same
invariant measure
\begin{equation}
\psi \left( \chi ,t\right) =\int \frac{d^{3}p}{\left( 2\pi \right)
^{3/2}\omega _{p}}\psi \left( p\right) e^{\frac{\mathrm{i}}{\hbar }\left(
-\omega _{p}t+p\chi \right) }=\left\langle \chi |\psi \left( t\right)
\right\rangle  \label{nova3}
\end{equation}%
Here we use the letter $\chi $ for position rather than $x$ for, as we shall
see, it cannot be interpreted as a position variable at all.

Then they imposed a set of physically reasonable assumptions on the
properties that localized states should obey and they found out that the
wave function describing a particle localized at position $x$ at $t=0$ is
given by%
\begin{equation}
\psi _{x}\left( p\right) =\sqrt{\frac{\omega _{p}}{\left( 2\pi \right) ^{3}}}%
e^{-\frac{\mathrm{i}}{\hbar }px}=\left\langle p|\psi _{x}\right\rangle
\label{nova2}
\end{equation}%
or, using (\ref{nova3}),%
\begin{equation}
\psi _{x}\left( \chi \right) =\int \frac{d^{3}p}{\left( 2\pi \right) ^{3}%
\sqrt{\omega _{p}}}e^{\frac{\mathrm{i}}{\hbar }p\left( \chi -x\right)
}=\left\langle \chi |\psi _{x}\right\rangle  \label{nova4}
\end{equation}%
which is in fact spread in space at distances of the order of the Compton
wavelength, and not a delta function. This has been sometimes erroneously
interpreted as non-localizability of particles in the Theory of Relativity.
In fact it is not possible to interpret simultaneously $\left\langle \chi
|\psi \right\rangle $ as the "wavefunction in position space" and $%
\left\langle p|\psi \right\rangle $ as the "wavefunction in momentum space"
as long as $\left\vert \chi \right\rangle $ and $\left\vert p\right\rangle $
are respectively the eigenkets of operators $X$ and $P$ which obey canonical
commutation relations among themselves, as we shall show.

Assuming that the sets of eigenkets of $X$ and $P$, respectively $\left\vert
x\right\rangle $ and $\left\vert p\right\rangle $, are complete, one can
write the following normalisations and partitions of unity%
\begin{eqnarray}
\left\langle x|y\right\rangle &=&g_{X}\left( x\right) \delta \left(
x-y\right) \\
\left\langle p|k\right\rangle &=&g_{P}\left( p\right) \delta \left(
p-k\right) \\
1 &=&\int \frac{dp}{g_{P}\left( p\right) }\left\vert p\right\rangle
\left\langle p\right\vert =\int \frac{dx}{g_{X}\left( x\right) }\left\vert
x\right\rangle \left\langle x\right\vert  \label{part}
\end{eqnarray}%
and the inner product is%
\begin{equation}
\left\langle \varphi |\psi \right\rangle =\int \frac{\left\langle \varphi
|x\right\rangle \left\langle x|\psi \right\rangle }{g_{X}\left( x\right) }%
dx=\int \frac{\varphi ^{\ast }\left( p\right) \psi \left( xp\right) }{%
g_{P}\left( p\right) }dp  \label{voi}
\end{equation}%
The point is that the canonical commutation relations impose further that%
\begin{equation}
\left\langle x|p\right\rangle =\sqrt{\frac{g_{X}\left( x\right) g_{P}\left(
p\right) }{2\pi \hbar }}\exp \left( \frac{\mathrm{i}}{\hbar }px\right)
\label{voo}
\end{equation}%
which, together with the partitions of unity (\ref{part}) leads to%
\begin{eqnarray}
\psi \left( p\right) &=&\sqrt{\frac{g_{P}\left( p\right) }{2\pi \hbar }}\int
\frac{dx}{\sqrt{g_{X}\left( x\right) }}\exp \left( -\frac{\mathrm{i}}{\hbar }%
px\right) \psi \left( x\right) \\
\psi \left( x\right) &=&\sqrt{\frac{g_{X}\left( x\right) }{2\pi \hbar }}\int
\frac{dp}{\sqrt{g_{P}\left( p\right) }}\exp \left( \frac{\mathrm{i}}{\hbar }%
px\right) \psi \left( p\right)  \label{nova}
\end{eqnarray}%
Now we see that eqs. (\ref{voi}) and (\ref{oito1}) imply $g_{P}\left(
p\right) \propto \omega _{p}$ while eqs. (\ref{nova}) and (\ref{nova3})
imply that $g_{P}\left( p\right) \propto \omega _{p}^{2}$. Therefore, if $%
\psi \left( p\right) $ is the wavefunction in momentum space, $\psi \left(
\chi \right) $ cannot be the wavefunction in position space. In which space
is it a representation, that is, which operator has $\left\vert \chi
\right\rangle $ as its eigenkets? Using the partition of unity (\ref{part})
one can compute%
\begin{equation}
\left\langle \chi _{1}|\chi _{2}\right\rangle =\int \frac{dp}{\omega _{p}}%
\left\langle \chi _{1}|p\right\rangle \left\langle p|\chi _{2}\right\rangle
=\int \frac{dp}{\omega _{p}}e^{\mathrm{i}p\left( \chi _{1}-\chi _{2}\right) }
\end{equation}%
The states $\left\vert \chi \right\rangle $ are not orthogonal to each other
and therefore cannot even correspond to the eigenstates of a hermitian
operator. The wavefunction $\psi \left( \chi \right) $ ia a representation
to which no observable is associated. Ref. \cite{pad} gives a clear
presentation of this point.

The eigenstates of the position operator are naturally the localized states
themselves, $\left\vert x\right\rangle \equiv \left\vert \psi
_{x}\right\rangle $, for which (\ref{nova2}) and (\ref{voo}) are compatible.
In this representation the wave function is given by (\ref{nova}) and not (%
\ref{nova3})%
\begin{equation}
\psi \left( x\right) =\sqrt{\frac{g_{X}\left( x\right) }{2\pi \hbar }}\int
\frac{dp}{\sqrt{\omega _{p}}}\exp \left( \frac{\mathrm{i}}{\hbar }px\right)
\psi \left( p\right)  \label{novo}
\end{equation}%
and localized states are indeed delta functions. Ref. \cite{pavsic} provides
us with a detailed exposition of the $\psi \left( \chi \right) $ and $\psi
\left( x\right) $ representations and how to transform between the two.

We should call the reader's attention to the reason why we could choose the
measure either as $\omega _{p}^{-1/2}$ or $\omega _{p}^{-1}$ in (\ref{fi})
but not in (\ref{novo}). The reason is that in (\ref{fi}) the choice of the
measure amounts only to a redefinition of the creation and annihilation
operators while in (\ref{novo}) the choice of the measure amounts to a
redefinition of the basis kets $\left\vert x\right\rangle $ - and they are
not free to choose, they have to be the eigenkets of an operator satisfying
canonical commutation relations with $P$. In summary, what Newton and Wigner
showed was that it is not possible to implement the canonical commutation
relations in a covariant manner. But that should not be surprising since the
canonical commutation relations are not covariant if we insist in
transforming the position operator with Lorentz transformations.

However, with the definition of position operator used here and with the
transformation law (\ref{pro2}) the canonical commutation relations become
covariant.

\section{\textbf{Antiparticles}}

We make a brief incursion into the complex Klein-Gordon field, described by
the Lagrangian%
\begin{equation}
L=\eta ^{\mu \nu }\partial _{\mu }\bar{\phi}\partial _{\nu }\phi -m^{2}\bar{%
\phi}\phi
\end{equation}%
in order to address the issue of antiparticles. The solution to the complex
Klein-Gordon equation involves two sets of creation and annihilation
operators%
\begin{eqnarray}
\phi \left( x\right) &=&\int \frac{d^{3}k}{\left( 2\pi \right) ^{3/2}}\sqrt{%
\frac{\hbar }{2\omega _{k}}}\left[ a_{k}e^{\mathrm{i}kx}+b_{k}^{+}e^{-%
\mathrm{i}kx}\right]  \label{kg10} \\
\pi \left( x\right) &=&\frac{\partial \bar{\phi}}{dt}\left( x\right) =\int
\frac{d^{3}k\mathrm{i}}{\left( 2\pi \right) ^{3/2}}\sqrt{\frac{\hbar \omega
_{k}}{2}}\left[ a_{k}^{+}e^{-\mathrm{i}kx}-b_{k}e^{\mathrm{i}kx}\right]
\end{eqnarray}%
The space of states becomes the cross product between the Fock spaces
generated by each set of creation operators%
\begin{equation}
\mathcal{F}=\mathcal{F}_{a}\otimes \mathcal{F}_{b}
\end{equation}

The number of particles operator splits into two parcels, each one of them
acting in each one of the factor spaces, and each one of them being
independently conserved%
\begin{eqnarray}
N &=&\int d^{3}ka_{k}^{+}a_{k}+\int d^{3}kb_{k}^{+}b_{k}=  \notag \\
&=&\frac{1}{\hbar }\int d^{3}x\frac{1}{2}\left( \bar{\phi}\mathcal{W}%
^{1/2}\phi +\mathrm{i}\left[ \bar{\phi}\bar{\pi}-\pi \phi \right] +\pi
\mathcal{W}^{1/2}\bar{\pi}\right) +  \notag \\
&&+\frac{1}{\hbar }\int d^{3}x\frac{1}{2}\left( \phi \mathcal{W}^{1/2}\bar{%
\phi}+\mathrm{i}\left[ \phi \pi -\bar{\pi}\bar{\phi}\right] +\bar{\pi}%
\mathcal{W}^{-1/2}\pi \right)
\end{eqnarray}%
Their difference is the charge operator%
\begin{equation}
Q=\frac{e}{2}\left[ \int d^{3}x\left[ \bar{\phi},\mathcal{W}^{1/2}\phi %
\right] +\left[ \pi ,\mathcal{W}^{1/2}\bar{\pi}\right] +\mathrm{i}\left\{
\bar{\phi},\bar{\pi}\right\} -\mathrm{i}\left\{ \pi ,\phi \right\} \right]
\end{equation}%
The energy, momentum, and position operators also split into two parcels
each one of them acting in each one of the factor spaces,%
\begin{eqnarray}
\vec{P} &=&\int d^{3}k\hbar \vec{k}a_{k}^{+}a_{k}+\int d^{3}k\hbar \vec{k}%
b_{k}^{+}b_{k}=  \notag \\
&=&\int d^{3}x\frac{1}{2}\left( -\vec{\nabla}\bar{\phi}\bar{\pi}-\pi \vec{%
\nabla}\phi +\mathrm{i}\left( \vec{\nabla}\bar{\phi}\mathcal{W}\phi +\vec{%
\nabla}\pi \mathcal{W}^{-1}\bar{\pi}\right) \right) +  \notag \\
&&+\int d^{3}x\frac{1}{2}\left( -\vec{\nabla}\phi \pi -\bar{\pi}\vec{\nabla}%
\bar{\phi}+\mathrm{i}\left( \vec{\nabla}\phi \mathcal{W}\bar{\phi}+\vec{%
\nabla}\bar{\pi}\mathcal{W}^{-1}\pi \right) \right)
\end{eqnarray}%
\begin{eqnarray}
H &=&\int d^{3}k\hbar \omega _{k}a_{k}^{+}a_{k}+\int d^{3}k\hbar \omega
_{k}b_{k}^{+}b_{k}=  \notag \\
&=&\int d^{3}x\frac{1}{2}\left( \pi \bar{\pi}+\vec{\nabla}\bar{\phi}\cdot
\vec{\nabla}\phi +m^{2}\bar{\phi}\phi +\mathrm{i}\left( \bar{\phi}\bar{\pi}%
-\pi \phi \right) \right) +  \notag \\
&&+\int d^{3}x\frac{1}{2}\left( \bar{\pi}\pi +\vec{\nabla}\phi \cdot \vec{%
\nabla}\bar{\phi}+m^{2}\phi \bar{\phi}+\mathrm{i}\left( \phi \pi -\bar{\pi}%
\bar{\phi}\right) \right)
\end{eqnarray}%
\begin{eqnarray}
\vec{X} &=&\mathrm{i}\int d^{3}ka_{k}^{+}\frac{\partial a_{k}}{\partial \vec{%
k}}+\mathrm{i}\int d^{3}kb_{k}^{+}\frac{\partial b_{k}}{\partial \vec{k}}=
\notag \\
&=&\frac{1}{\hbar }\int d^{3}x\vec{x}\frac{1}{2}\left( \left[ \left(
\mathcal{W}^{1/2}\bar{\phi}\right) -\mathrm{i}\left( \mathcal{W}^{-1/2}\pi
\right) \right] \left[ \left( \mathcal{W}^{1/2}\phi \right) +\mathrm{i}%
\left( \mathcal{W}^{-1/2}\bar{\pi}\right) \right] \right) +  \notag \\
&&+\frac{1}{\hbar }\int d^{3}x\vec{x}\frac{1}{2}\left( \left[ \left(
\mathcal{W}^{1/2}\phi \right) -\mathrm{i}\left( \mathcal{W}^{-1/2}\bar{\pi}%
\right) \right] \left[ \left( \mathcal{W}^{1/2}\bar{\phi}\right) +\mathrm{i}%
\left( \mathcal{W}^{-1/2}\pi \right) \right] \right)
\end{eqnarray}

It should be pointed out that antiparticles do not disappear in the
transition from Quantum Field Theory to Quantum mechanics: all subspaces
with fixed numbers of particles show up, be it one particle, two particles,
one antiparticle, one particle plus one antiparticle, etc. One last
observation is that we do not agree with the conclusion withdrawn in \cite%
{pad} that "... the consistent description of the NRQM limit of QFT requires
us to work with a pair of fields, corresponding to a particle and its
antiparticle". In fact we have done it here for the real Klein-Gordon field,
and in this could also have been done in \cite{pad} too if in his section 6
the author had defined $\phi \left( x\right) =A\left( x\right) +A^{+}\left(
x\right) $ rather than introducing a new field $B\left( x\right) $ and
defining $\phi \left( x\right) =A\left( x\right) +B^{+}\left( x\right) $.

\section*{Conclusions}

In this work, we have analysed the non-interacting Klein-Gordon field as a
tool to construct the algebra of operators acting on the Hilbert spaces of
Quantum Mechanics for systems of $N$ identical particles from the field
operators acting in the Fock space of Quantum Field Theory. This is achieved
by relating the position and momentum quantum operators with the field
operators. Future work is required to understand whether this recipe could
also be applied or at least adapted to interacting fields. The main point is
that all polynomial operators on the positions and momenta acting on any of
the $N$-particle subspaces of Fock space can be constructed out of the field
operators. The position operators so constructed turn out to be the
Newton-Wigner operators. Under Lorentz transformations, they do not
transform as tensors, rather in a manner that preserves the canonical
commutation relations.

As a by-product, we showed that regardless of the Fermi-Dirac or
Bose-Einstein statistics of field, the position and momentum operators obey
commutation relations.

Finally we showed that, contrary to what is claimed in \cite{pad}, the
transition from Quantum Field Theory to Quantum Mechanics can be obtained
without resource to antiparticles. However, if the field theory describes
particles and antiparticles, then two single particle quantum mechanical
systems can be extracted from it: one with one particle and one with one
antiparticle; as well as all possible combinations of numbers of particles
and antiparticles.

\end{document}